\documentclass[conference, a4paper,10pt]{IEEEtran}
\usepackage[T1]{fontenc}% optional T1 font encoding
\pdfoutput=1
\IEEEoverridecommandlockouts
\usepackage[font=footnotesize,caption=false]{subfig} % preload with correct options...
\usepackage{tikz}
\usepackage{cite}
\usepackage{tikz}
\usepackage{cite}
\usepackage{amsmath,amssymb,amsfonts}
\usepackage{soul}
\usepackage{ulem}
\usepackage{enumitem}
\usepackage{tikz}
\allowdisplaybreaks
\usepackage{pgfplots}
\pgfplotsset{width=14cm,compat=1.9}
\usepackage[noend,ruled,algo2e]{algorithm2e}
\usetikzlibrary{arrows}
\usetikzlibrary{arrows.meta}
\usetikzlibrary{automata, positioning}
\usepackage{algorithmic}
\usepackage{graphicx}
\usepackage{textcomp}

\usepackage{xcolor}
\usepackage{booktabs}
\usepackage[short]{optidef}
\usepackage{mathtools}
\usepackage{glossaries}
\usepackage{subfig}
\usepackage{algorithm}
\usepackage{bbm}
\usepackage{multirow}

\usepackage{mathtools,amsthm}
\newtheorem{theorem}{Theorem}

\newtheorem{lemma}{Lemma}

\def\BibTeX{{\rm B\kern-.05em{\sc i\kern-.025em b}\kern-.08em
    T\kern-.1667em\lower.7ex\hbox{E}\kern-.125emX}}

\definecolor{color3}{HTML}{FFD700}
\definecolor{color2}{HTML}{EA5F94}
\definecolor{color1}{HTML}{9D02D7}
\definecolor{color0}{HTML}{0000FF}

\newcommand{\pp}[1]{\textcolor{red}{#1}}

\newcommand{\indctr}{\mathbbm{1}}
\newcommand{\remove}[1]{}

\usepackage{xcolor,cancel}

\title{On the Regret of Online Coded Caching}
\remove{\author{Anonymous authors}}
\author{\IEEEauthorblockN{Anupam Nayak}
\IEEEauthorblockA{\textit{Dept. of Electrical Engg.} \\
\textit{IIT Bombay}\\
Mumbai, India \\
anupam@ee.iitb.ac.in}
\and
\IEEEauthorblockN{Sheel F Shah}
\IEEEauthorblockA{\textit{Dept. of Electrical Engg.} \\
\textit{IIT Bombay}\\
Mumbai, India \\
sheelfshah@gmail.com}  
\and
\IEEEauthorblockN{Nikhil Karamchandani}
\IEEEauthorblockA{\textit{Dept. of Electrical Engg.} \\
\textit{IIT Bombay}\\
Mumbai, India \\
nikhilk@ee.iitb.ac.in}  
}

\begin{document}

\maketitle

\begin{abstract}
We consider the widely studied problem of `coded caching under non-uniform requests' where users independently request files according to some underlying popularity distribution in each slot. This work is a first step towards analyzing this framework through the lens of online learning. We consider the case where the underlying request
distribution is apriori unknown and propose an online policy
as well as study its regret with respect to an oracle which
knows the underlying distribution and employs a well-known
order-optimal placement and coded delivery strategy. We also
bound the switching cost of this strategy and also discuss a
lower bound on the regret of any online scheme in a restricted but natural class of policies.
\end{abstract}

\section{Introduction}
The last few decades have witnessed an unprecedented growth of demand for high-definition content over the internet. The resulting increased load on the underlying communication networks has been mitigated by the widespread adoption of content delivery networks, which deploy storage devices or caches across large geographical regions. These caches are then used to pre-fetch popular content in the off-peak hours. This cached content can then be used to reduce network traffic during peak hours when users make the most requests. 

Caching has a rich history, see for example \cite{wessels2001web} and references therein. More recently, an information-theoretic study of caching networks began with the seminal work of Maddah-Ali and Niesen \cite{maddah2014fundamental, maddah2014decentralized}. This has led to a large amount of literature, under the moniker `coded caching', studying various aspects of such systems, including non-uniform content popularity \cite{ccarb, %deng2022fundamental, 
hachem2017coded,
% niesen2016coded
hachem2014multi, hachem2015effect, ji2017order}, network topology \cite{karamchandani2016hierarchical, hachem2017coded}, security and privacy \cite{gurjarpadhye2022fundamental, ravindrakumar2017private} etc. Broadly speaking, these works propose content placement and delivery schemes and then compare their performance with that of the information-theoretic optimal, often showing a gap of at most a constant multiplicative factor independent of the system size. 

In particular, our work considers the framework of `coded caching under non-uniform requests' \cite{ccarb, hachem2017coded, niesen2016coded} where at each time, users request files according to an underlying request distribution. The aforementioned works assume that the request distribution is known and then design order-optimal placement and delivery schemes. In this work, we consider a scenario where the request distribution is apriori unknown, and the system has to adapt its content placement and delivery scheme over time based solely on the requests observed at the users. We measure the performance of any such \textit{online policy} using the metric of \textit{regret}, which is standard in the online learning literature. For any given time horizon $T$, the regret of a scheme considers the cumulative additive gap with respect to an oracle which knows the underlying distribution beforehand. We propose a natural online policy for the coded caching framework and demonstrate that its regret is bounded by an instance-dependent constant, and thus does not scale with the time horizon. Given that the online policy changes the content placement across time, we also consider the switching cost of our policy in terms of the number of timesteps where the cached content needs to be updated and are again able to show a constant upper bound for this cost. Finally, we also prove a lower bound on the regret of any online policy in a certain restricted class. 

The study of caching systems in the framework of online learning has been pursued recently by other works as well. Most of these focus on the case of a single cache. Inspired by the recent advances in online convex optimization \cite{cohen2015following, abernethy2014online, zinkevich2003online}, several online caching policies have been proposed including Online Gradient Ascent \cite{paschos2019learning, paschos2020online}, Online Mirror Descent \cite{salem2021no}, and Follow the Perturbed Leader (FTPL) \cite{bhattacharjee2020fundamental, paria2021texttt, zarin2022regret}. These works consider the case of \textit{adversarial requests} and demonstrate that an order-optimal regret of $\Theta(\sqrt{T})$ can be achieved. On the other hand, there have also been some works which consider \textit{stochastic requests} \cite{bura2021learning, zarin2022regret} and establish tighter bounds on the expected regret. Our work is similar in spirit and is a first step to studying the popular coded caching framework from the lens of online learning. 

The rest of the paper is organized as follows. We describe the problem setup formally in Section~\ref{sec:sysmodel}. In Section~\ref{sec:results1}, we describe our proposed online caching policy and establish upper bounds on its regret as well as switching cost. Section~\ref{sec:results2} considers a restricted but natural class of online policies and derives a lower bound on the regret of any policy in this class. Numerical results are presented in section \ref{sec:numexp}. We conclude with a short discussion in Section~\ref{sec:conclusion}. Due to lack of space, the proofs of some of the theorems and lemmas stated are presented in \cite{full_paper}.

%\section{Background Work}\label{sec:bgw}

\section{System Model}\label{sec:sysmodel}
\begin{figure}[ht]
    \centerline{\scalebox{.8}{\includegraphics{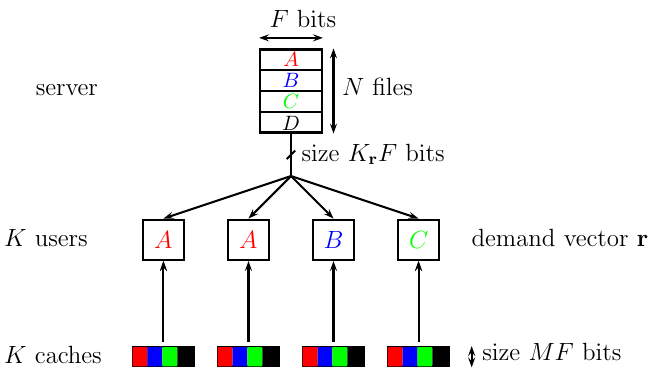}}}
    \caption{System setup for the coded caching problem}
    \label{Fig:System}

\end{figure}
Consider the caching system depicted in Figure~\ref{Fig:System}. It consists of a single server hosting $N$ files $W_1, W_2, \ldots, W_N$, each of size $F$ bits. The server is connected via an error-free broadcast link to $K$ users, each equipped with a local cache of size $MF$ bits. Consider time to be divided into slots and in each slot $t$, the system operates in two phases: a \textit{placement} phase followed by a \textit{delivery} phase. The placement phase happens during the low network traffic hours, and in this phase, the cache at each user $k$ is populated with content related to the $N$ files, say $Z_k(t)$. Note that the content stored in each cache can vary over time, possibly as a function of prior requests. Next, during the delivery phase, each user $k$ requests the file with index $r_k^t \in \{1,2,\ldots, N\}$ which is generated according to an underlying probability mass function (pmf) $\mathbf{p} = (p_1, p_2, \ldots, p_N)$. Without loss of generality, we assume $p_1 \geq p_2 \geq \cdots \geq p_N$. The requests are assumed to be independent and identically distributed (i.i.d.) across time and users. Furthermore, the request distribution $\mathbf{p}$ is apriori unknown to the caching system. The request profile $\mathbf{r^t} = (r_1^t, r_2^t, \ldots, r_K^t)$ is forwarded to the server, and depending on the requests and the cached contents, the server transmits a message $X(t)$ of size $K_{\mathbf{r^t}}F$ bits on the shared link to the users. Each user $k$ should be able to recover its requested file $W_{r_k^t}$ using $X(t)$ and its cached content $Z_k(t)$. For any feasible online (placement + delivery) policy $\pi$ and over any given time horizon $T$, the overall (normalized) expected server transmission size is given by $\mathcal{K}^{\pi}(T) = \mathbb{E}_{\mathbf{p}}[\sum_{t=1}^{T}  K_{\mathbf{r^t}}]$, where the expectation is with respect to the underlying request distribution $\mathbf{p}$.   

In the same spirit as standard online learning problems, we will compare the performance of any online policy with that of a (static) oracle which knows the underlying request distribution $\mathbf{p}$ and uses a fixed placement over the time horizon $T$, chosen to minimize the expected rate. This problem has been studied in the literature under the moniker `coded caching under non-uniform requests' \cite{ccarb, hachem2017coded, niesen2016coded}. While characterizing the optimal scheme under a general request distribution is very challenging, there have been schemes proposed that are order-optimal, i.e., their expected server transmission rate is within a constant multiplicative\footnote{some results have an additional constant additive gap to the optimal also.} factor of the information-theoretic optimal.
In particular, we will consider as benchmark\footnote{Using an approximation algorithm as a benchmark for defining regret is common in the online learning literature when finding the optimal policy is intractable, see for example \cite{azar2022alpha, li2021online}.} the scheme proposed in \cite{ccarb}.

%Some intuition to justify using this scheme for benchmarking and coded caching\cite{maddah2014fundamental}, in general, can be seen from the toy example below. 
\begin{table*}
\begin{center}
\begin{tabular}{|c|c|c|c|c|c|c|} 
 \hline
$C_1$ & $C_2$ & $M_{AA}$ & $M_{AB}$ & $M_{BA}$ & $M_{BB}$ & avg rate \\  
 \hline
 $[A_1, A_2]$ & $[A_1, A_2]$ & $\times$ & $B$& $B$ & $B$ & 0.75  \\ 
 \hline
$[A_1, A_2]$ & $[B_1, B_2]$ & $A$ & $\times$ & $B\oplus A$ & $B$ & 0.75 \\
 \hline
$[A_1, B_1]$ & $[A_2, B_2]$ & $A_2\oplus A_1$ & $A_2\oplus B_1$ & $B_2\oplus A_1$ & $B_2\oplus B_1$ & 0.5\\
 \hline
\end{tabular}
\end{center}
\caption{Summary of transmission rates for various request patterns and cache contents}
\vspace{-9mm}
\label{table:tx}
\end{table*}

We illustrate the main idea of coded caching \cite{maddah2014fundamental} via a simple toy example. Say we have  $N = 2$ files $A, B$, and $K = 2$ users with caches $C_1$ and $C_2$ of size $M = 1$ each. Assume that the files are equally popular and thus the request distribution is $[0.5,0.5]$ for both users. This makes all the possible request patterns $AA$, $AB$, $BA$, and $BB$ equally likely. Let $A_1, A_2$ denote two halves of file $A$. As shown in Table \ref{table:tx}, when $[A_1, A_2]$ is stored in both the caches, the server broadcast transmission will be $B$ for all requests except $AA$ when no transmission is required ($\times$), bringing the average transmission size to 0.75. Similarly storing $[A_1,A_2]$ in $C_1$ and $[B_1,B_2]$ in $C_2$ will require server transmissions to be as per Table \ref{table:tx}. Note that when $BA$ is requested, the server transmits the coded message $B\oplus A$. User 1 can retrieve $B$ by performing an XOR operation on the message with the file $A$ from its cache. Similarly, user 2 can retrieve $A$. 
%The length of $B\oplus A$ is the same as any of the two files, and the message length still happens to be 1, 
The average transmission size for this storage profile also turns out to be 0.75. 
%will be observed whenever we store any set of complete files in the caches. 
Finally, we consider the case where $[A_1,B_1]$ is stored in $C_1$ and $[A_2, B_2]$ is stored in $C_2$. Here, each of the transmissions is of size 0.5, giving a smaller average rate than above, while using coded server transmissions to serve all user requests. %For e.g., for the request $AB$, the transmission will be $A_2\oplus B_1$ where user 1 retrieves $A$ jointly from $A_1$ present in the cache and $A_2$ obtained by performing XOR operation on the transmitted message with $B_1$ from its cache. 

The above illustrated ideas were generalized in \cite{maddah2014fundamental} to any collection of parameters $N, K, M$. 
%Similar division of files into subfiles for a reduced average rate will require increased coordination among users as the number of users and files becomes large. 
Furthermore, a decentralized coded caching scheme was proposed in \cite{maddah2014decentralized} which enabled each user to randomly and independently sample $M / N$ bits from each file and then use a coded delivery scheme as before.  However, both \cite{maddah2014fundamental, maddah2014decentralized} assume all files to be equally popular, which does not always hold in practice. For non-uniform popularity distributions, we will consider as benchmark the order-optimal scheme proposed in \cite{ccarb}, which uses the same placement and delivery policy as \cite{maddah2014decentralized} but only on files that have popularity higher than a particular system-dependent 
threshold. More details are provided below.
%The idea is to not waste cache space by populating it with unpopular files and instead, transmitting these files fully using the server when they are (infrequently) requested.  

%for a cache size of $M$ and $N$ equally popular files 
The placement and delivery phases for the scheme in \cite{ccarb} are defined as follows: 
\begin{enumerate}[label=\roman*.]
%\color{red}
\item \textit{Placement}: All files whose underlying request probability is less than the threshold $1 / (MK)$ are not stored in any cache. For the remaining, say $N_1$, most popular files, the disparity in their request probabilities is ignored and the placement policy proposed under the `decentralized coded caching' scheme for uniformly popular files  in \cite{maddah2014decentralized} is used.  
\item \textit{Delivery}: For unpopular files that have not been cached, the server serves the requests directly. For the $N_1$ popular files stored in the caches, the corresponding coded delivery policy from \cite{maddah2014decentralized} is used. 
\end{enumerate}
For any underlying request distribution $\mathbf{p}$, the achievable normalized expected server transmission size, say $K_o$, in any slot for the scheme\footnote{A more detailed description of the placement and delivery scheme from \cite{ccarb} is provided in Appendix \ref{app:scheme}} above, denoted henceforth by $\pi^\star$,  is shown in \cite{ccarb} to be at most 
\begin{equation}
\label{Eqn:RateOracle}
K_o \leq \left[\frac{N_1}{M}-1\right]_{+}+\min\left\{\sum_{i > N_1} K p_i , \frac{N-N_1}{[M-N_1]_{+}}-1 \right\}.
\end{equation}
Furthermore, the expected transmission rate for any caching scheme $\pi$ is lower bounded \cite{ccarb} by
\begin{equation}
\label{Eqn:RateOraclelb}
K_{\pi} \geq \max\left\{\frac{1}{29}\left[\frac{N_1}{M}-1\right]_{+}, \frac{1}{58}\left[\sum_{i > N_1} K p_i -2\right]_{+} \right\} \overset{\Delta}{=} K_{lb}
\end{equation}
The upper and lower bounds are shown in \cite{ccarb} to be within a constant multiplicative and additive gap, independent of the system size.
%These two can be related as per \cite{ccarb} as
%\begin{equation}
%K_{ub} \leq 87 K_{lb}+2
%\end{equation}

For any online policy $\pi$, we will define the \textit{regret} over a time horizon $T$ as the difference between its expected transmission size and that of the oracle scheme described above, i.e., 
\begin{equation}
\label{Eqn:Regret}
R_\pi(T) = \mathcal{K}^{\pi}(T) - T . K_o = \sum_{t=1}^{T}\mathbb{E}[K_{\pi}(t) - K_{o}] . 
\end{equation}
This work aims to design an online policy with provably small regret w.r.t the oracle, without knowing the underlying distribution. In fact, as we demonstrate in Section~\ref{sec:results1}, a scheme which essentially has the same placement and delivery policy as $\pi^\star$ above, while using the empirical request distribution in each slot instead of the true underlying one for performing cache placement, has constant regret which does not scale with the time horizon $T$. We also provide a lower bound on the regret of any online policy in a restricted class. 

Note that the placement scheme for an online policy can change over time slots and updating the cache contents will also incur some cost. For our proposed policy, we also provide an upper bound on the \textit{switching cost} in terms of the overall number of time slots over any horizon $T$, say $S(T)$, where cache contents are updated.  

For any integer $n \ge 1$, we will denote the set $\{1,2,\ldots,n\}$ by $[n]$. Also, for integers $n_2 > n_1 \ge 1$, let $[n_1:n_2]$ denote the subset $\{n_1, n_1+1, \ldots, n_2\}$. 
\remove{
We begin with considering a Content Distribution Network with a single server connected to $K$ users through an errorless broadcast channel for the distribution of requested files. There are $N$ files, each of size $F$ bits, on the server, and during time slot $t$, the request of user $k$ is denoted by $d_k(t) \in [N]$. These requests are independent and identically distributed according to a popularity distribution $(p_1,p_2 \cdots p_N)$ across users and time slots (with $\sum\limits_{i=1}^{N}p_i = 1$). Without loss of generality, we assume $p_1 \geq p_2 \geq \cdots \geq p_N$. Each user has a local cache of size $MF$ bits.

User requests are revealed at the beginning of each time slot. The server sees the requests  $d_k(t) \ \forall \ k \in [K]$ and transmits an appropriate message using the shared broadcast link such that each user can retrieve the requested file using the local memory bits in the cache and the broadcast message. (\textcolor{red}{we could also write this as calculating the file $d_k(t)$ as a function of the local cache and the broadcasted message}). At the end of this slot, the server can update the cache for every user based on the information available until time $t$ (request history). The cost/rate incurred during this slot is equal to the length of the transmitted message $K_{\pi}(t)$, and it is assumed that the cache updates occur between time slots, where the rate incurred by the server is not significant. 

In \cite{ccarb} the authors come up with an order optimal coding policy for this setting, when the file popularity distribution is known. The first part of our work analyses the performance of a mean tracking algorithm (when this distribution is unknown) in comparison to the policy suggested in \cite{ccarb}. This involves using the estimated popularities $(\hat{p}_1(t)\cdots \hat{p}_N(t))$ within the delivery and placement phase of the algorithm at time $t$. Here $\hat{p}_i(t)$ simply denotes the fraction of the $K \cdot t$ requests upto (and excluding) time $t$ that were for file $i$. 

The Regret of this policy for a horizon $T$ is defined as the expected total number of extra bits that needed to be transmitted during the delivery phase of each slot compared to an Oracle that knows the file popularity distribution and caches using the policy proposed in \cite{ccarb}. Note that expected rate for the Oracle is the same across time steps (i.e. $K_o$).
\[R(T) = \sum_{t=1}^{T}\mathbb{E}[K_{\pi}(t) - K_{o}(t)] = \sum_{t=1}^{T}\mathbb{E}[K_{\pi}(t)] - K_{o}\]
Further, one can note that the cached content of each user under oracle's policy will be constant across time steps, while the cache can potentially change when working with the tracking policy. This will involve additional transmissions during the placement phase for cache updates. These extra transmissions can be accounted for by associating a switching regret with them.
\[S(T) = \sum_{t=1}^{T}\mathbb{E}[D \mathbb{I}(\text{Switch}(t))  N_s(t)]\]. Where $N_s(t)$ is the number of bits involved in the switching. In Section \ref{sec:results1}, we discuss the performance of the tracking policy and upper bound the regrets by a constant. Next, in section \ref{sec:results2}, we analyze the performance of the tracking approach for the optimal policy for coded delivery with uncoded placement proposed in \cite{2opt} for the special case if $N$ files with just two users.
}
\section{Online scheme and Achievable Regret}\label{sec:results1}

In this section, we propose an online placement and delivery policy $\mathcal{P}$ which does not apriori know the underlying request distribution $\mathbf{p}$. 
\begin{itemize}
    \item Calculate the estimated popularity distribution $\{\mathbf{\hat{p}}^t\}_{t \ge 1}$ from requests seen until $t-1$.
    \item Perform placement and delivery for the estimated
    "popular" files as per \cite{maddah2014decentralized}.
\end{itemize}Our policy tracks the empirical request probability distribution $\{\mathbf{\hat{p}}^t\}_{t \ge 1}$ based on the actual requests observed, and then assuming this to be the actual distribution, in each round performs caching according to the placement and delivery policy for the oracle scheme $\pi^\star$ as described before. That is, we use the placement and delivery policy in \cite{maddah2014decentralized} to cache all files with empirical popularity at least $1 / (KM)$ ("popular" files), while not storing the rest of the files in any cache and serving any requests for these files directly via the server. Our first result characterizes the regret of this online policy $\mathcal{P}$ with respect to the oracle policy $\pi^\star$, as defined in  \eqref{Eqn:Regret}.  

\remove{
We propose a policy that keeps track of the empirical probabilities of each file being requested and uses these empirical means to perform caching using the strategy described in \cite{ccarb} (i.e. use the placement policy in \cite{maddah2014decentralized} to cache all files with empirical popularity at least $\frac{1}{KM}$, and not store rest of the files in any cache). Since there are no empirical means available in the first time slot, our policy performs decentralized caching as in \cite{maddah2014decentralized} for all files.
}
%\subsection{Upper Bounds}
\begin{theorem}\label{theo:UB1}
The regret of our proposed online caching policy $\mathcal{P}$ can be upper bounded as 
\begin{equation*}
    R_{\mathcal{P}}(T) \leq \min\{A, B\} + \underbrace{\frac{N}{M} - 1 - K_o}_{\text{first step regret}}
    %  \frac{(2 + \sum\limits_{i=1}^N \Delta_i)(K-K_o)}{K\Delta^2} + \dfrac{K(1 - M/N)}{1 + KM/N} - K_o
\end{equation*}
where $K_o$ is the expected (per slot) server transmission size for the oracle policy $\pi^\star$, $\Delta_i = \left|p_i - \frac{1}{KM}\right|$, $\Delta = \min_i \Delta_i$,  and 
$$
A = \frac{(2 + \sum\limits_{i=1}^N \Delta_i)(K-K_o)}{K\Delta^2}, B = \frac{4(K-K_o)}{K\Delta^2}. 
$$
\end{theorem}
\remove{
\begin{theorem}\label{theo:UB2}
\begin{equation*}
    R_\pi(T) \leq \frac{4(K-K_o)}{K\Delta^2} + \dfrac{K(1 - M/N)}{1 + KM/N} - K_o
\end{equation*}
\end{theorem}
\pp{can write both as one theorem with rhs as the min of the two}
}
The above result demonstrates that the proposed scheme $\mathcal{P}$ achieves a constant regret with respect to the oracle, which does not scale with the time horizon $T$. 
%\subsection{Proofs of theorems 1, 2}
%
\begin{proof}
Let $[N_1] = \{1,2,\ldots, N_1\}$ denote the set of file indices for which the underlying request probability is greater than the threshold $1 / (KM)$ (popular set). Then, we have 
\begin{align*}
    \Delta &= \min_i \Delta_i = \min\bigg\{p_{N_1}-\frac{1}{KM}, \frac{1}{KM}-p_{N_1+1}\bigg\}
    \end{align*}
   % \Delta_i &= \left|p_i - \frac{1}{KM}\right|\\
For each file $W_i$, define the empirical request probability at the start of time slot $t$ as 
\begin{align*}
    \hat{p}_i^t &= \begin{cases}
        \dfrac{\sum\limits_{j=1}^{t-1} \sum\limits_{k=1}^K \mathbbm{1}\{r_k^j = i\}}{(t-1)K}, &\quad t \geq 2 ,\\
        \dfrac{1}{N}, &\quad t = 1 .
    \end{cases}
\end{align*}
Note that in each slot, the oracle policy $\pi^\star$ caches the subset of files with indices in $[N_1]$ according to the decentralized coded caching in \cite{maddah2014decentralized}, whereas our policy applies the same scheme to the subset $\{ i \in [N] : {\hat{p}_i^{t}} \geq \frac{1}{KM}\}$ at time $t > 1$. For $t = 1$, all the $N$ files are included in the subset chosen to cache. 

\subsubsection{Upper bounding using Chernoff bounds}
For each $t \geq 1$, let $\mathcal{G}^t$ denote the event that $\hat{p}_i^{t} > p_i - \Delta_i \ \forall \ i \in [N_1]$ and $\hat{p}_i^{t} < p_i + \Delta_i \ \forall \ i \in [N_1+1:N]$. Under this event, we have that both the oracle and the proposed policy will cache the same files in slot $t$ since $\forall \; j \in [N_1]$ we have 
{$\hat{p}_j^{t} > p_j-\Delta_j \geq 1 / (KM)$ and $\forall \; k \in [N_1+1:N]$ we get $\hat{p}_k^{t} < p_k + \Delta_k \leq 1 / (KM)$.}

%{\color{red}$\hat{p}_j^{t-1} \geq p_j-\Delta \geq p_{N_1} - \Delta \geq \frac{1}{KM}$ and  $\forall \; k \in [N_1+1:N]$ we get $\hat{p}_k^{t-1} \leq p_k + \Delta \leq p_{N_1+1} + \Delta \leq \frac{1}{KM}$.}
%
\begin{lemma}
Let $\Bar{\mathcal{G}^t}$ denote the complement of $\mathcal{G}^t$
\begin{equation}
    \mathbb{P}(\Bar{\mathcal{G}^t}) \leq \sum_{i=1}^{N} \exp \left( -\frac{\delta_i^2(t-1)Kp_i}{2+\delta_i} \right)
\end{equation}  
where $\delta_i = \Delta_i / p_i$
and total regret is bounded as
\begin{equation}
\label{Eqn:B1}
 R_{\mathcal{P}}(T) \leq \frac{(2 + \sum_{i=1}^N \Delta_i)(K-K_o)}{K\Delta^2} + \frac{N}{M} - 1 - K_o .
\end{equation}
\end{lemma}
The proof of this lemma can be found in Appendix~\ref{App:l1} of \cite{full_paper}
\subsubsection{Upper bounding using the DKW inequality} We now present another way to upper bound the regret of the proposed scheme $\mathcal{P}$. Consider the event $\mathcal{H}^t: \max_i |\hat{p}_i^{t}-p_i| \le \Delta$. It follows from \cite[Lemma 2]{bura2021learning}, which is an application of the Dvoretzky–Kiefer–Wolfowitz (DKW) inequality \cite{dvoretzky1956asymptotic}, that 
\begin{equation}
\label{Eqn:DKW}
    \mathbb{P}(\Bar{\mathcal{H}^t}) \leq 2e^{-(t-1)K\Delta^2/2} .
\end{equation}
Under the event $\mathcal{H}^t$, we have that both the oracle policy and the proposed policy will take the same action since $\forall \; j \in [N_1]$ we have $\hat{p}_j^{t} \geq p_j-\Delta \geq p_{N_1} - \Delta \geq \frac{1}{KM}$ and  $\forall \; k \in [N_1+1:N]$ we get $\hat{p}_k^{t} \leq p_k + \Delta \leq p_{N_1+1} + \Delta \leq \frac{1}{KM}$.
\remove{
Lemma 2 in [\ref{learning_to_cache_caching_to_learn}] gives us \pp{(or re-derive if needed?)}
\begin{equation}
    \mathbb{P}\bigg(\max_{i\in N} |\hat{p}_i^t-p_i| > \epsilon\bigg) \leq 2e^{-tK\epsilon^2/2}
\end{equation}
}
As before, we can bound the regret after the first time slot can be bounded as 
\begin{align*}
    \label{eqn:regret_breakdown}
    &\sum_{t=2}^{T}(K-K_o)\cdot\mathbb{P}(\Bar{\mathcal{H}^t}) \overset{(a)}{\leq} 2(K-K_{o})\sum_{t=2}^{T}e^{-(t-1)K\Delta^2/2}\\
    &\leq 2(K-K_{o})\sum_{t=2}^{\infty}e^{-(t-1)K\Delta^2/2}
\leq \frac{4(K-K_o)}{K\Delta^2} 
\end{align*}
where $(a)$ follows from \eqref{Eqn:DKW}. 
%Where (\ref{eqn:regret_breakdown}) follows from the fact that under $G^t$, our algorithm has the same expected rate as the Oracle, and outside it, our algorithm incurs a rate of at most $K$ (which is equivalent to sending each of the requested files as is).\\
%The rate for the first time, assuming uniform caching, is at most $K(1 - M/N)\dfrac{1}{1 + KM/N}$. Since $\frac{e^{-x}}{1-e^{-x}}\leq\frac{1}{x} \; \forall x>0$ we have
Thus the overall regret for $\mathcal{P}$ can be bounded as 
\begin{equation}
\label{Eqn:Bound2}
    R_{\mathcal{P}}(T) \leq \frac{4(K-K_o)}{K\Delta^2} + \frac{N}{M} - 1 - K_o .
\end{equation}
Combining \eqref{Eqn:B1} and \eqref{Eqn:Bound2} completes the proof. In order to express the bound in terms of only the problem parameters, one may lower bound $K_o$ by $K_{lb}$
\end{proof}

\subsection{Switching cost}
Note that the placement scheme for an online policy can change over time slots, and updating the cache contents will also incur some cost. For our proposed policy, we now provide an upper bound on the expected switching cost in terms of the overall number of time slots over any horizon $T$, say $S(T)$, where
cache contents are updated. In fact, we show that the expected switching cost is bounded above by an instance-dependent constant and does not scale with the horizon $T$. 
\remove{
In order to account for the transmissions involved in fetching the files to update the cache before every slot, one can include a switching cost and define the total regret as defined as
\begin{equation}
R(t) = \sum_{t=1}^{T}(\mathbb{E}[K_{\pi}(t)] - K_{o}) + \sum_{t=1}^{T}D \mathbb{E}[\mathbb{I}(\text{Switch}(t))]
\end{equation}
A cost of $D$ is associated with each switch, the indicator evaluates to 1 if a transmission is to be made before slot $t$ in order to update the local caches.
}
\begin{theorem}\label{theo:switch1}
%For regret defined as $\sum\limits_{t=1}^{T}D \mathbb{I}(\text{Switch}(t))$
    For any horizon $T$, the expected switching cost for our proposed policy $\mathcal{P}$ can be upper-bounded as
    \[ S(T) \le 1 + \sum_{i \in [N_1]}\frac{1+\beta_i}{2K\Delta_i^2} + \sum_{i \in [N] \backslash [N_1] } \frac{\alpha_i+1}{2K\Delta_i^2}  \]
 where $[N_1] = \{1,2,\ldots, N_1\}$ denotes the set of file indices for which the underlying request probability is greater than the threshold $1 / (KM)$, and 
 \begin{align*}
     \alpha_i &= e^{2\Delta_i^2} \sum\limits_{j=0}^{\lfloor\frac{1}{M}\rfloor} {{K}\choose{j}}p_i^j(1-p_i)^{K-j} , \\
     \beta_i &= e^{2\Delta_i^2}\sum\limits_{j=\lfloor\frac{1}{M}\rfloor+1}^{K} {{K}\choose{j}}p_i^j(1-p_i)^{K-j} .
 \end{align*}
\end{theorem}
%$\mathcal{P}$ is the true "popular set" with files whose true popularities $p_i\geq \frac{1}{KM}$
Note that a cache update happens at time $t$ when the estimated "popular set" changes. This happens either when an existing file is removed from the popular set (event $\mathcal{A}^t$), or a new file gets included in the popular set (event $\mathcal{B}^t$). Note that these events are not disjoint.
 \begin{lemma}
      The probability of events $\mathcal{A}^t$ and $\mathcal{B}^t$ can be bounded as
$$
   \mathbb{P}(\mathcal{A}^t) \le \sum_{i \in [N_1]}
     {e^{-2K(t-1)\Delta_i^2}} +  \sum_{i \in [N] \backslash [N_1]}\alpha_i
     e^{-2K(t-1)\Delta_i^2} 
$$        
 \end{lemma}

\[ \mathbb{P}(\mathcal{B}^t) \le \sum_{i \in [N_1]}\beta_i
     e^{-2K(t-1)\Delta_i^2}+\sum_{i \in [N]\backslash [N_1]}
     e^{-2K(t-1)\Delta_i^2}.\]
The proof of the above lemma can be found in Appendix~\ref{app:lemmaswitch} in \cite{full_paper}. The proof of Theorem \ref{theo:switch1} can be found in Appendix~\ref{app:thm2} in \cite{full_paper}.

\section{Lower bound}
\label{sec:results2}
Our next result is a lower bound on the expected regret of any online policy that does not apriori know the underlying request distribution. We restrict ourselves to a class, say $\mathcal{C}$, of policies that, in each round $t$, pick some subset of file indices $S^t$, and use the 'decentralized coded caching' placement and delivery policy in \cite{maddah2014decentralized} to uniformly store files corresponding to this subset and serve their requests. If $|S^t| \geq M$, then the files outside this subset are not stored in any cache, and requests for such files are directly served by the server in the delivery phase. If $|S^t|<M$, the cache memory remaining after completely storing all files in $S^t$ is used to perform uniform decentralized coded caching for the remaining files. Note that our proposed policy $\mathcal{P}$ from the previous section also belongs to this class $\mathcal{C}$. 

For any policy $\pi \in \mathcal{C}$ and a given  system instance $\mathcal{I} = (N, K, M, \mathbf{p})$, we consider the expected rate\footnote{The rate expression used here is an approximation (in fact, an upper bound) for the actual expected rate of the scheme $\pi$, as follows from \cite{ccarb, maddah2014decentralized}. The true expression is cumbersome, and hence we use this approximation to facilitate our analysis. However, a similar line of proof will apply there as well.} incurred in round $t$ by $\pi$ to be  \begin{equation}\label{eqn:apprxrate}
K_{S_\pi^t}(\mathcal{I}) = \begin{cases}
    \frac{|S_\pi^t|}{M} -1  + \sum_{i \not\in S_\pi^t} Kp_i & |S_\pi^t| > M\\
    \frac{N - |S_\pi^t|}{M - |S_\pi^t|} - 1 & |S_\pi^t| \leq M 
\end{cases}
\end{equation}
We will assume that $N$ is even for brevity in our proof. 

\begin{theorem}\label{theo:LB2}
    Consider any caching system with parameters $N, M, K$ such that $N / K < M < N/2$. Fix any $0 < a < b$ with $\frac{1}{a} + \frac{1}{b} = \frac{1}{N}$ and $\frac{2}{b} < \frac{1}{KM} < \frac{2}{a}$. Then for any policy $\pi$ in the class $\mathcal{C}$ described above, there exists a horizon $T$ and an underlying request distribution (parameterized by $a, b$) such that
    \begin{align*}
        R_\pi(T) \geq \frac{K(\frac{1}{KM} - \frac{2}{b})}{8\left(\frac{1}{a}-\frac{1}{b}\right) \log\left(\frac{b}{a}\right)} . 
    \end{align*}
\end{theorem}
\begin{proof}
We choose $a < b$ such that $\frac{1}{a} + \frac{1}{b} = \frac{1}{N}$ and $\frac{1}{KM} \in (\frac{2}{b}, \frac{2}{a})$; such a pair will always exist for $KM > N$.

%\textcolor{blue}{Sidenote: Such $a, b$ will always exist. Eg. let $b= 4KM \implies \frac{2}{b}=\frac{1}{2KM}<\frac{1}{KM}$ and $\frac{2}{a} = \frac{2}{N} - \frac{1}{2KM} > \frac{2}{KM} - \frac{1}{2KM} > \frac{1}{KM}$}

Next, we define
\begin{align*}
    \mathbf{p} &:= \left\{p_1 = \frac{2}{a}\cdots p_{N/2} = \frac{2}{a},p_{N/2+1} = \frac{2}{b}\cdots p_N = \frac{2}{b}\right\}\\
    \mathbf{q} &:= \left\{q_1 = \frac{2}{b}\cdots q_{N/2} = \frac{2}{b},q_{N/2+1} = \frac{2}{a}\cdots q_N = \frac{2}{a}\right\}
\end{align*}
and create two caching system instances $\mathcal{I}_p$, $\mathcal{I}_q$ with parameters $N, K, M$ and underlying request distributions $\mathbf{p}$ and $\mathbf{q}$ respectively.

For an instance $\mathcal{I}$,  define $S_1 = \left[\frac{N}{2}\right]$ and $S_2 = \left[\frac{N}{2}+1:N\right]$. %Thus, we have  $S_g(\mathcal{I}_p) = S_1, S_b(\mathcal{I}_p) = S_2$ and $S_g(\mathcal{I}_q) = S_2, S_b(\mathcal{I}_q) = S_1$.
A subset $S \subseteq [N]$ is said to be $S_1$-dominant if $\frac{|S \cap S_1|}{|S|} \geq \frac{1}{2}$, i.e., the majority of elements in $S$ belong to $S_1$. Otherwise, it is said to be $S_2$-dominant (equivalently, $\frac{|S \cap S_2|}{|S|} > \frac{1}{2}$).  We say that a caching subset $S$ is "bad" if it is $S_2$-dominant when the underlying distribution is $\mathbf{p}$ or if it is $S_1$-dominant when the underlying distribution is $\mathbf{q}$. 

%\frac{|S \cap S_b|}{|S|} \geq \frac{1}{2}$, i.e.,
%define $S_b(\mathcal{I}) = \arg\min_i p_i$ and $S_g(\mathcal{I}) = \arg\max_i p_i$. Also
Note that the oracle which knows the underlying request distribution will select files with a request probability larger than $1/KM$ for caching. That is, it will select $S_1$ for caching if the instance was $\mathcal{I}_{p}$ and $S_2$ if the instance was $\mathcal{I}_{q}$. Hence, the expected rate for the oracle in each slot is given by $K_o = \frac{N}{2M} - 1 + \frac{KN}{b}$ for both the instances $\mathcal{I}_p, \mathcal{I}_q$ from \eqref{eqn:apprxrate}.

Now lemma \ref{lemma:khalf_lemma} provides a lower bound on the regret incurred by any policy in a timeslot when it selects a "bad" subset of files to cache. 
\end{proof}
\begin{lemma}\label{lemma:khalf_lemma}
For any policy $\pi \in \mathcal{C}$ and either instance $\mathcal{I}_p$ or $\mathcal{I}_q$, let $S$ be the set of files selected by the policy for caching  for instance at some timeslot $t$. Let $K_S$ denote the expected rate incurred in the slot. If $S$ is "bad", then
\begin{equation*}
    K_S - K_o \geq \Gamma, \mbox{ where } \Gamma \overset{\Delta}{=} \frac{NK}{2}(\frac{1}{KM} - \frac{2}{b})
\end{equation*}
\end{lemma}
This lemma is proved in Appendix \ref{app:lemma1} in \cite{full_paper}

Now for any policy $\pi \in \mathcal{C}$ and some time horizon $T$, consider event $\mathcal{A}$ that the subsets of files selected for caching by $\pi$ at different timeslots, denoted by $S_\pi^1, S_\pi^2, \cdots, S_\pi^T$, are $S_2$-dominant for at least half the slots. Formally,
\begin{equation*}
    \mathcal{A} = \left\{ \sum_{t=1}^T \indctr \left\{ \frac{|S_\pi^t \cap S_2|}{|S_\pi^t|} > \frac{1}{2} \right\} \geq \frac{T}{2}\right\}
\end{equation*}
Thus we have
\begin{equation*}
    \overline{\mathcal{A}} = \left\{ \sum_{t=1}^T \indctr \left\{ \frac{|S_\pi^t \cap S_1|}{|S_\pi^t|} \geq \frac{1}{2} \right\} \geq \frac{T}{2}\right\}
\end{equation*}
For instance $\mathcal{I}_p$, occurrence of event $\mathcal{A}$ for a policy $\pi$ implies that $S_\pi^t$ is "bad" for at least $T/2$ rounds. Similarly, for instance $\mathcal{I}_q$, occurrence of event $\overline{\mathcal{A}}$ for a policy $\pi$ implies that $S_\pi^t$ is "bad" for at least $T/2$ rounds.

Let $R_\pi^p(T)$ and $R_\pi^q(T)$ denote the regret incurred by $\pi$ on the instances $\mathcal{I}_p$ and $\mathcal{I}_q$ with underlying request distributions $\mathbf{p}$ 
 and $\mathbf{q}$ respectively. 
 %Then we have
 %the first instance with file popularities $\mathbf{q}$ and similarly $R_2 (T)$ is the regret incurred by $\pi$ on the second instance with file popularities $\mathbf{q'}$. 
\remove{
\begin{align*}
    \mathbb{E}[R(T)] &= \mathbb{P}(\mathcal{A})\mathbb{E}[R(T)|\mathcal{A}] +\mathbb{P}(\overline{\mathcal{A}})\mathbb{E}[R(T)|\overline{\mathcal{A}}]\\
    &\geq \mathbb{P}(\mathcal{A})\mathbb{E}[R(T)|\mathcal{A}]
\end{align*}
}
Under the event $\mathcal{A}$ for instance $\mathcal{I}_p$, the cache is "bad" for at least $\frac{T}{2}$ steps. From Lemma~\ref{lemma:khalf_lemma}, the difference in the rate incurred by the policy $\pi$ and the oracle is at least $\Gamma$ in each of these $\frac{T}{2}$ steps. A similar argument holds for the instance $\mathcal{I}_p$ under the event $\overline{\mathcal{A}}$. 
%Therefore $\mathbb{E}[R(T)|A]\geq \frac{T}{2}(K_{\frac{1}{2}}-K_{o}) $ the minimum regret incurred under instance $\mathcal{I}_1$ is
Thus, we have 
\begin{equation}
\label{Eqn:Regret2Instances}
R_\pi^p(T) \ge \frac{T\Gamma}{2} \mathbb{P}_{\mathbf{p}}^{\pi}(\mathcal{A}), \quad R_\pi^q(T) \ge \frac{T\Gamma}{2} \mathbb{P}_{\mathbf{q}}^{\pi}(\overline{\mathcal{A}}) .
\end{equation}
\remove{
\begin{align*}
    R_1 &\geq \frac{T}{2}(K_{\frac{1}{2}}-K_{o})\mathbb{P}_{q}^{\pi}(A)\\
    R_2 &\geq \frac{T}{2}(K_{\frac{1}{2}}-K_{o})\mathbb{P}_{q'}^{\pi}(\overline{A})
\end{align*}
The second equation comes from extending the same argument to $\mathcal{I}_2$ and the event $\overline{A}$.

Now using the alternative version of the Bretagnolle-Huber inequality stated in \pp{[\ref{bhi}]} we have
\begin{equation*}
    \mathbb{P}_{p}^{\pi}(\mathcal{A}) + \mathbb{P}_{q}^{\pi}(\overline{\mathcal{A}}) \geq \frac{1}{2}\exp(-D_{KL}(\mathbb{P}_p^\pi||\mathbb{P}_q^\pi))
\end{equation*}
where $\mathbb{P}_p^\pi$ and $\mathbb{P}_q^\pi$ denote the probability measures on the sequence of requests and cached files by the (possibly random) algorithm under their respective instances. 
}
Thus we obtain:
\begin{align*}
    &\max\{R_\pi^p(T),R_\pi^q(T)\} \\
    &\geq \frac{1}{2}(R_\pi^p(T)+R_\pi^q(T)) \geq \frac{T\Gamma}{2}(\mathbb{P}_{\mathbf{p}}^{\pi}(\mathcal{A}))+\mathbb{P}_{\mathbf{q}}^{\pi}(\overline{\mathcal{A}}))\\
    & \overset{(a)}{\geq} \frac{T\Gamma}{4}\exp(-D_{KL}(\mathbb{P}_p^\pi||\mathbb{P}_q^\pi))\\
    &\overset{(b)}{=} \frac{T\Gamma}{4} \exp\left(-TN\left(\frac{1}{a}-\frac{1}{b}\right) \log\left(\frac{b}{a}\right)\right)
\end{align*}
where $(a)$ follows from the Bretagnolle-Huber inequality \cite[Theorem 14.2]{torlat}; and $(b)$ follows from Lemma~\ref{lemma:ddl} in Appendix~\ref{app:ddl} in \cite{full_paper} which is based on a divergence decomposition result. The theorem statement then follows by choosing the horizon $T$ which maximizes the expression on the right hand side of the above inequality.  
\remove{
\begin{proof}
\begin{equation*}
    \geq \frac{T\Gamma}{4}\exp(-D_{KL}(\mathbb{P}_p^\pi||\mathbb{P}_q^\pi))
    \label{div 1}
\end{equation*}
From the divergence calculation in the subsequent subsection we can write \ref{div 1} as
\begin{equation*}
     = \frac{T}{4}(K_{\frac{1}{2}}-K_o) \exp\left(-TN\left(\frac{1}{a}-\frac{1}{b}\right) \log\left(\frac{b}{a}\right)\right)
\end{equation*}

Note that the function attains its maximum at $T_{max} = \frac{1}{N\left(\frac{1}{a}-\frac{1}{b}\right)\log\left(\frac{b}{a}\right)}$ and then falls off, But since the minimum expected regret over the set of policies operating with a horizon of $T_1$ should be less than the minimum expected regret for the set of algorithms operating with a horizon of $T_2$ for $T_2>T_1$ we have the lower bound equal to 
\[
    = \begin{cases}
        \frac{T\Gamma}{4} e^{\left(-T/T_{\max}\right)} & T\leq T_{\max}\\
        \frac{T_{\max}\Gamma}{4} & T>T_{\max} \\
    \end{cases} \]

Plugging in the \pp{$K_{1/2}$} expression obtained above gives the desired expression for a large enough $T$.
\end{proof}}

\section{Numerical experiments}
\label{sec:numexp}
In this section, we compare the performance of our policy using two benchmarks.
%To the best of our knowledge, there is no work that proposes policies for online multi-user caching.Thus, we use the following benchmarks:
\begin{enumerate}
    \item \textbf{Uniform coded caching}:  This policy uses the same placement and delivery scheme used by us described in \cite{maddah2014decentralized} for the cached files . However, it does not perform thresholding of any sort and caches all the files uniformly ($N/M$ fraction of each file).  
    \item \textbf{Uncoded LFU}: This policy estimates the popularities after each time step and caches the top $M$ files with the highest popularity estimates. Cached files are directly served, incurring 0 rate, and uncached files are entirely transmitted by the server.
\end{enumerate}

All the simulations are performed using estimated file popularities from the Movielens 1M dataset \cite{ML1m_paper}, which contains $\sim$ 1 million ratings from 6040 users on 3706 movies. Popularities are assigned using the fraction of ratings for a movie out of all ratings. The file requests were generated by sampling from this distribution and the regret for each policy was calculated as the difference of its rate from the oracle rate\footnote{We use the upper bound on the achievable oracle rate (described in section \ref{sec:sysmodel}) as this is known to be a good approximate for the actual rate for large $K$ ($K$=400 was used) as shown in \cite{ccarb}.}.

\begin{figure}
    \centering 
    \subfloat[K = 400, M = 10]{\includegraphics[width=0.9\linewidth]{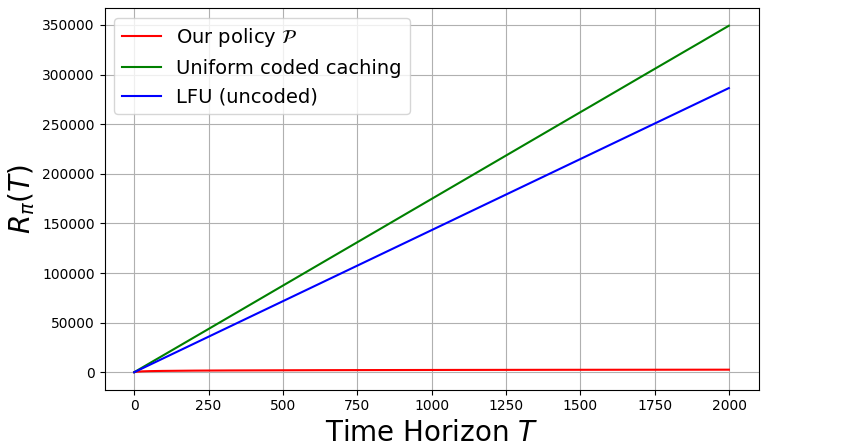}\label{fig:plot2}}\hskip1ex
    \subfloat[K = 400, M = 100]{\includegraphics[width=0.9\linewidth]{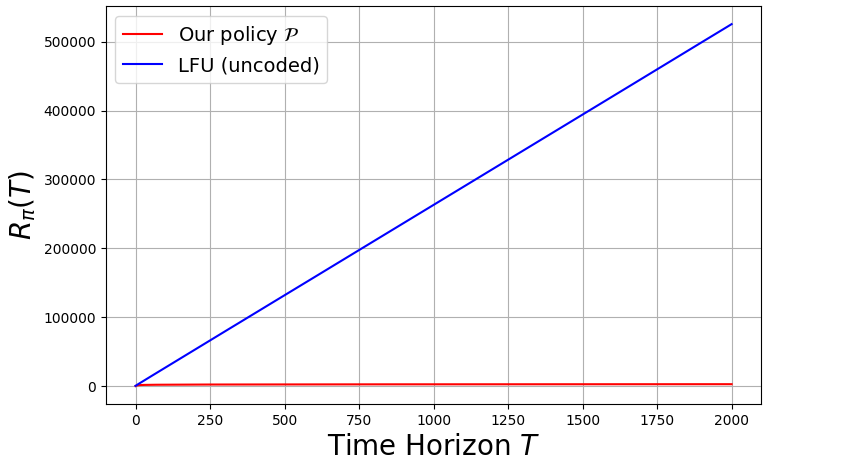}\label{fig:plot3}}\hskip1ex
    \subfloat[K = 400, M = 100]{\includegraphics[width=0.9\linewidth]{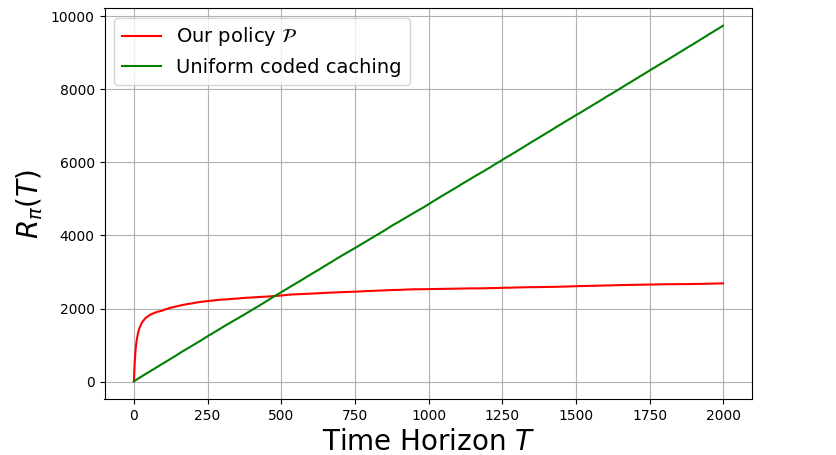}\label{fig:plot4}}\hskip1ex
    \caption{
    A comparison of benchmark policies against the proposed policy $\mathcal{P}$. In (a), our policy significantly outperforms Uniform Coded Caching \cite{maddah2014decentralized} and uncoded LFU for small $M$. In (b), our policy is shown to outperform uncoded LFU for large $M$. Finally, in (c), we see that for large $M$, the regret for Uniform Coded Caching grows linearly with the time horizon $T$, whereas the regret of our policy $\mathcal{P}$ approaches a constant. This empirically confirms the bounds described in Theorem \ref{theo:UB1}.
    }
    \vspace{-10pt}
\end{figure}

As seen in Figs.~\ref{fig:plot2}-\ref{fig:plot4}, our proposed policy, $\mathcal{P}$, outperforms the benchmarks. The uncoded LFU method performs significantly worse for large $M$, which emphasizes the superiority of coded schemes over uncoded ones. Moreover, the expected one-step regret is constant for both, the LFU and Uniform coding schemes. Hence, their regrets are seen to grow linearly with $T$. On the other hand, the regret for our policy $\mathcal{P}$ approaches a constant. This provides empirical verification for Theorem~\ref{theo:UB1}.

\section{Conclusions and Future Work}\label{sec:conclusion}
This work is a first step towards analyzing the widely studied coded caching framework through the lens of online learning. We consider the case where the underlying request distribution is apriori unknown and propose an online policy as well as study its regret with respect to an oracle that knows the underlying distribution and employs a well-known order-optimal placement and coded delivery strategy. We also bound the switching cost of this strategy and also discuss a lower bound on the regret of any online scheme in a restricted class. 

There are several avenues for future work. Firstly, there are more complicated coded caching schemes other than the threshold-based policy considered here which are known to have better performance and can serve as the oracle benchmark for regret analysis. Secondly, while we restrict attention here to the case of stochastic requests from a stationary request distribution, there are several other interesting scenarios including a temporally varying request distribution and adversarial requests.

% \appendices
% \section{Lemma} 

% For a binomial random variable we have
% \[\sqrt{np}-\frac{1-p}{2\sqrt{np}} \leq \mathbb{E}[\sqrt{X}]\leq\sqrt{np}\]
% Note that both inequalities are tight for $np\gg 1$ i.e. for for large $np$ we have $\mathbb{E}[\sqrt{X}] \approx \sqrt{np}$ .
% The right side of the lemma follows directly from Jensen's inequality. Note that for some $y>0$ and $\mu> 0$ we have 
% \[(y-\sqrt{\mu})^2(y+2\sqrt{\mu}) = y^3 - 3\mu y + 2 \mu^{3/2} \geq 0\]
% let $y=\sqrt{x}$ thus
% \[3\mu\sqrt{x}-x^{3/2} \leq 2\mu^{3/2}\]
% \[3\mu x-x^2 \leq 2\mu^{3/2}\sqrt{x}\]
% \[2\mu^2 + \mu(x-\mu) - (x-\mu)^2 \leq 2\mu^{3/2}\sqrt{x}\]
% Thus dividing throughout by $2\mu^{3/2}$ we have
% \[\sqrt{\mu} + \frac{x-\mu}{2\sqrt{\mu}} - \frac{(x-\mu)^2}{2\mu^{3/2}} \leq \sqrt{x} \]
% taking expectation on both sides and using $\mu = np$ and variance $np(1-p)$ we have 
% \[\sqrt{np}-\frac{1-p}{2\sqrt{np}} \leq \mathbb{E}[\sqrt{X}]\]

\bibliographystyle{IEEEtran}
\bibliography{references}
\section{Appendix}\label{sec:app}
\subsection{Placement and Delivery Scheme}\label{app:scheme}
Here we describe the placement and delivery scheme from \cite{ccarb}. The scheme essentially involves the use of the placement and delivery scheme described in \cite{maddah2014decentralized} for a subset of files. This subset includes all the files with popularity $\geq\frac{1}{KM}$. The rest of the files are "unpopular" and are fully delivered by the server whenever requested.
\subsubsection{Placement}
In the cache placement stage at any time step $t$, once the selection of files $S$ to be cached is determined, each user independently selects $\min\left\{|F|, \frac{M|F|}{|S|}\right\}$ bits randomly from the chosen file to store in their cache. Note that the set $S$ for this policy consists of files for which the observed request probability until time $t$ exceeds $\frac{1}{KM}$. The collection of files cached by the oracle remains constant as it knows the true request probability distribution, enabling it to select files that exceed the threshold of $\frac{1}{KM}$. If the number of files crossing this threshold is less than $M$, we store all the bits from the first $M$ files and divide the remaining storage amongst the rest of the files uniformly.
\subsubsection{Delivery}
Let $U_1$ denote the set of users that request files from the cached set $S$, $|U_1| = K_4$
\begin{itemize}
    \item For every subset $s \in U_1$ with $|s|\neq 0$ transmit $\bigoplus_{k\in s} V_{k,s\setminus\{k\}}$.
    \item For every request from $U_1^c$ directly transmit the whole file ($|F|$ bits).
\end{itemize}
Here  $ V_{k,s\setminus\{k\}}$ denotes all the bits that are requested by user $k$, are present in the cache of all users in $s$ and that are not stored in the caches of any other user
in $U_1\setminus s $.  $\bigoplus_{k\in s} V_{k,s\setminus\{k\}}$. denotes a XOR operation across all the users in $s$, i.e, a XOR across  all the sets $V_{k,s\setminus\{k\}}$. Note that every user in $s$ will be able to retrieve the bits designated to it from $\bigoplus_{k\in s} V_{k,s\setminus\{k\}}$ by repeatedly performing XOR operations using the bits present in its own cache. By construction, every bit requested by any user in $U_1$ has to be present in one of the sets $V_{k,s\setminus\{k\}}$.

\subsection{Upper Bounds : Proof of Lemma 1}
\label{App:l1}
Let $\Bar{\mathcal{E}}$ denote the complement of event $\mathcal{E}$. Then, we have 
\begin{align}
\nonumber
&\mathbb{P}(\Bar{\mathcal{G}^t}) 
%\nonumber
= \mathbb{P}\left(\bigcup_{i=1}^{N_1} \{p_i-\hat{p}_i^{t} \ge  \Delta_i\} \cup \bigcup_{i=N_1+1}^{N} \{\hat{p}_i^{t} - p_i \ge \Delta_i\}\right)\\
\nonumber
    &\leq \sum_{i=1}^{N_1} \mathbb{P}( p_i-\hat{p}_i^{t} \ge \Delta_i) + \sum_{i=N_1+1}^{N} \mathbb{P}( \hat{p}_i^{t} - p_i \ge \Delta_i)\\
 \nonumber
    &= \sum_{i=1}^{N_1} \mathbb{P} \left(\hat{p}_i^{t} \le p_i(1 - \frac{\Delta_i}{p_i}) \right) + \sum_{i=N_1+1}^{N} \mathbb{P} \left(\hat{p}_i^{t} \ge p_i(1+\frac{\Delta_i}{p_i}) \right)\\
 \nonumber
    &\overset{(a)}{\leq} \sum_{i=1}^{N_1} e^{ -\frac{\delta_i^2(t-1)Kp_i}{2}} + \sum_{i=N_1+1}^{N} e^{ -\frac{\delta_i^2(t-1)Kp_i}{2+\delta_i}}\\
    \label{eqn:UpperBound}
    &\leq \sum_{i=1}^{N} \exp \left( -\frac{\delta_i^2(t-1)Kp_i}{2+\delta_i} \right)
\end{align}
where $\delta_i = \Delta_i / p_i$ and $(a)$ follows from the Chernoff bound.  
Next, we will evaluate the regret of our proposed policy. Recall that for the first time slot, we essentially assume all files to be equally popular and use the decentralized coded caching scheme from \cite{maddah2014decentralized} over the set of all files. From \cite{maddah2014decentralized}, the expected server transmission size for this scheme is at most $N/M$, and hence the regret in the first slot is at most $N/M - K_o$. In any subsequent slot $t$, there will be no regret if the event $\mathcal{G}^t$ occurs since the proposed policy would be taking the same actions as the oracle policy. If $\mathcal{G}^t$ does not occur, then the regret in the slot will be at most $K - K_o$. We have
\begin{align*}
    \label{eqn:regret_breakdown}
  %\sum_{t=2}^{T}\mathbb{P}(G^t)(0) +
  &\sum_{t=2}^{T}(K-K_o) \cdot \mathbb{P}(\Bar{\mathcal{G}^t})\\
    &\overset{(a)}{\leq} (K-K_{o})\sum_{t=2}^{T}\sum_{i=1}^{N} \exp \left( -\frac{\delta_i^2}{2+\delta_i}(t-1)Kp_i \right)\\
   % &= (K-K_{o})\sum_{i=1}^{N}\sum_{t=2}^{T} \exp \left( -\frac{\delta_i^2}{2+\delta_i}(t-1)Kp_i \right)\\
    &\leq (K-K_{o})\sum_{i=1}^{N}\sum_{t=2}^{\infty} \exp \left( -\frac{\delta_i^2}{2+\delta_i}(t-1)Kp_i \right)\\
    &= (K-K_{o})\sum_{i=1}^{N}\left(\frac{\exp \left(-\frac{\delta_i^2}{2+\delta_i}Kp_i \right)}{1-\exp \left(-\frac{\delta_i^2}{2+\delta_i}Kp_i \right)}\right)\\
    &\overset{(b)}{\leq} (K-K_{o})\sum_{i=1}^{N}\left(\frac{2 + \delta_i}{\delta_i^2 K p_i}\right) \\
    &= (K-K_{o})\sum_{i=1}^{N}\left(\frac{2p_i + \Delta_i}{\Delta_i^2 K}\right)
    \leq  (K-K_{o}) \cdot \frac{2 + \sum_{i=1}^N \Delta_i}{\Delta^2 K}
\end{align*}
where $(a)$ follows from \eqref{eqn:UpperBound}; and $(b)$ follows since for each $x > 0$, $e^{-x} / (1-e^{-x})\leq 1 / x$. Thus the overall regret of our policy $\mathcal{P}$  is upper bounded as 
\begin{equation}
\label{Eqn:Bound1}
 R_{\mathcal{P}}(T) \leq \frac{(2 + \sum_{i=1}^N \Delta_i)(K-K_o)}{K\Delta^2} + \frac{N}{M} - 1 - K_o .
\end{equation}

\subsection{Proof of Lemma 2}\label{app:lemmaswitch}
\begin{proof}
\begin{align*}
   \mathbb{P}(\mathcal{A}^t) &= \mathbb{P}\left(\cup_{i=1}^{N} \{i \in \mathcal{N}_1^{t-1} , i \notin  \mathcal{N}_1^{t}\}\right)\\
   %&\leq \sum_{i \in N_1(t)}\mathbb{P}(i^{th}\text{ file jumps out})\\
     & \le \sum_{i \in N}\mathbb{P}\left(\{i \in \mathcal{N}_1^{t-1} , i \notin  \mathcal{N}_1^{t}\}\right)\\
    & = \sum_{i \in N}\mathbb{P}\left(\hat{p}_i(t)\leq\frac{1}{KM}, \hat{p}_i(t-1)> \frac{1}{KM}\right)\\
    & = \sum_{i \in [N_1]}\mathbb{P}\left(\hat{p}_i(t)\leq\frac{1}{KM}, \hat{p}_i(t-1)> \frac{1}{KM}\right) \\
     &+ \sum_{i \in [N] \backslash [N_1]}\mathbb{P}\left(\hat{p}_i(t)\leq\frac{1}{KM}, \hat{p}_i(t-1)> \frac{1}{KM}\right)
\end{align*}
where recall that $[N_1]$ denotes the set of popular files corresponding to the true underlying distribution. For the popular files, we have
\begin{align*}
     & \sum_{i \in [N_1]}\mathbb{P}\left(\hat{p}_i(t)\leq\frac{1}{KM}, \hat{p}_i(t-1)> \frac{1}{KM}\right)\\
    % & =\sum_{i \in \mathcal{P}}\mathbb{P}\left(\hat{p}_i(t-1)> \frac{1}{KM}\middle|\hat{p}_i(t)\leq\frac{1}{KM}\right)
    % \mathbb{P}\left(\hat{p}_i(t)<\frac{1}{KM}\right)\\
     %& \leq \sum_{i \in \mathcal{P}}\mathbb{P}\left(\hat{p}_i(t-1)\leq \frac{1}{KM}+\frac{1}{t}\right) \mathbb{P}\left(\hat{p}_i(t)<\frac{1}{KM}\right)\\
     %& \leq \sum_{i \in \mathcal{P}}\mathbb{P}\left(\hat p_i-{p}_i(t-1)\geq \Delta_i-\frac{1}{t}\right) \mathbb{P}\left(\hat{p}_i(t)<\frac{1}{KM}\right)
     & \leq \sum_{i \in [N_1]}
     \mathbb{P}\left(\hat{p}_i(t)\le \frac{1}{KM}\right) \leq  \sum_{i \in [N_1]}
     e^{-2K(t-1)\Delta_i^2}
\end{align*}
where the last step follows from Hoeffding's inequality. For the unpopular files, we have 
\begin{align*}
     & \sum_{i \in [N] \backslash [N_1]}\mathbb{P}\left(\hat{p}_i(t)\leq\frac{1}{KM}, \hat{p}_i(t-1)> \frac{1}{KM}\right)\\
     & =\sum_{i \in [N] \backslash [N_1]}\mathbb{P}\left(\hat{p}_i(t)< \frac{1}{KM}\middle|\hat{p}_i(t-1)\geq\frac{1} {KM}\right) \times \\ 
     & \hspace{.75in} \mathbb{P}\left(\hat{p}_i(t-1)\geq\frac{1}{KM}\right)\\
     %& \leq \sum_{i \in \mathcal{P}^c}\mathbb{P}\left(\hat{p}_i(t)\leq \frac{1}{KM}-\frac{1}{t}\right) \mathbb{P}\left(\hat{p}_i(t-1)>\frac{1}{KM}\right)\\
     %& \leq \sum_{i \in \mathcal{P}^c}\mathbb{P}\left(\hat {p}_i(t)-p_i\geq \Delta_i-\frac{1}{t}\right) \mathbb{P}\left(\hat{p}_i(t-1)>\frac{1}{KM}\right)
     & \leq \sum_{i \in [N] \backslash [N_1]}\mathbb{P}\left(\hat{p}_i(t)< \frac{1}{KM}\middle|\hat{p}_i(t-1)\geq\frac{1} {KM}\right) {e^{-2K(t-2)\Delta_i^2}}
     \\
     &\leq \sum_{i \in [N] \backslash [N_1]}\mathbb{P}\left(\sum\limits_{k=1}^K \mathbbm{1}\{r_k^t = i\}  < 1 /M  \right)e^{-2K(t-2)\Delta_i^2}\\
     &=  \sum_{i \in [N] \backslash [N_1]}\alpha_i
     e^{-2K(t-1)\Delta_i^2} .
\end{align*}
Thus overall, the probability of event $\mathcal{A}^t$ can be bounded as
$$
   \mathbb{P}(\mathcal{A}^t) \le \sum_{i \in [N_1]}
     {e^{-2K(t-1)\Delta_i^2}} +  \sum_{i \in [N] \backslash [N_1]}\alpha_i
     e^{-2K(t-1)\Delta_i^2} 
$$  
\remove{
Note the last step here comes from the argument that there need to be less than $K.\frac{1}{KM}$ requests for the file $i$ at time $t-1$ to push it below the threshold given that it was below the threshold at $t-1$. The probability of this happening is 

\[\sum\limits_{j=0}^{\lfloor\frac{1}{M}\rfloor} {{K}\choose{j}}p_i^j(1-p_i)^{K-j}.\]
}
Similarly, the probability of event $\mathcal{B}^t$ can be bounded as
\[ \mathbb{P}(\mathcal{B}^t) \le \sum_{i \in [N_1]}\beta_i
     e^{-2K(t-1)\Delta_i^2}+\sum_{i \in [N]\backslash [N_1]}
     e^{-2K(t-1)\Delta_i^2}.\]
\end{proof}
\subsection{Proof of Theorem 2}
\label{app:thm2}
\begin{proof}

 Recall that in the first slot, our policy includes all files in the popular set and thereafter only those files for which the empirical request probability is larger than the threshold $1/(KM)$. So we incur at most cost $1$ at $t = 2$ and consider $t > 2$ hereafter. Let the total number of file requests for the file $i$ before time $t$ be denoted by $L_i^{t-1}$. Thus the empirical popularity $\hat{p}_i^t$ for file $i$ will be $\frac{L_i^{t-1}}{K(t-1)}$. Depending on the number of requests observed for the file in time slot $t$, the updated empirical popularity $\hat{p}_i^{t+1}$ can range from $\frac{L_i^{t-1}}{Kt}$ to $\frac{L_i^{t-1}+K}{Kt}$. It is easy to verify that we always have $|\hat{p}_i(t+1)-\hat{p}_i(t)|\leq \frac{1}{t}$.
 We will denote the estimated popular set of files at time $t$ by $\mathcal{N}_1^t$. 
Thus, we have the following upper bound on the expected switch cost of our proposed policy $\mathcal{P}$:
\begin{align*}
%&\sum_{t=1}^{T}D \mathbb{E}[\mathbb{I}(\text{Switch}(t))]\\
 %& \leq D + \sum_{t=2}^{T}D\mathbb{P}({I}(\text{Switch}(t)=1)\\
 & 1 + \sum_{t=2}^{T}(\mathbb{P}(\mathcal{A}^t)+\mathbb{P}(\mathcal{B}^t))\\
 & \le 1 + \sum_{t=2}^{T}\left(\sum_{i \in [N_1]}
     e^{-2K(t-1)\Delta_i^2} + \sum_{i \in [N] \backslash [N_1]}\alpha_i
     e^{-2K(t-1)\Delta_i^2}\right)
 \\ &+ \sum_{t=2}^{T}\left(\sum_{i \in [N]}\beta_i
     e^{-2K(t-1)\Delta_i^2}+\sum_{i \in [N] \backslash [N_1]}
     e^{-2K(t-1)\Delta_i^2} \right)\\
 & \leq 1 + \left(\sum_{i \in [N_1]}
     \frac{e^{-2K\Delta_i^2}}{1-e^{-2K\Delta_i^2}} + \sum_{i \in [N] \backslash [N_1]}\alpha_i
     \frac{e^{-2K\Delta_i^2}}{1-e^{-2K\Delta_i^2}}\right)
 \\ &+ \left(\sum_{i \in [N_1]}\beta_i
     \frac{e^{-2K\Delta_i^2}}{1-e^{-2K\Delta_i^2}}+\sum_{i \in [N] \backslash [N_1]}
     \frac{e^{-2K\Delta_i^2}}{1-e^{-2K\Delta_i^2}} \right) \\   
      & \leq 1 + \left(\sum_{i \in [N_1]}
     \frac{1}{2K\Delta_i^2} + \sum_{i \in [N] \backslash [N_1]}\alpha_i
     \frac{1}{2K\Delta_i^2}\right)
 \\ &+ \left(\sum_{i \in [N_1]}\beta_i
     \frac{1}{2K\Delta_i^2}+\sum_{i \in [N] \backslash [N_1]}
     \frac{1}{2K\Delta_i^2} \right) \\
     &= 1 + \sum_{i \in [N_1]}\frac{1+\beta_i}{2K\Delta_i^2} + \sum_{i \in [N] \backslash [N_1] } \frac{\alpha_i+1}{2K\Delta_i^2} .
\end{align*}
\end{proof}

\subsection{Proof of Lemma 3}\label{app:lemma1}
\begin{proof}
Consider instance $\mathcal{I}_p$ and the case where $|S|>M$. Then, 
\begin{align*}
    K_S(\mathcal{I}_p) &= \frac{|S|}{M} + K\sum_{i \not\in S}p_i-1\\
    &\overset{(a)}{\geq} \frac{|S|}{M} + K \frac{2}{b} \left\lfloor \frac{N-|S|}{2} \right\rfloor + K \frac{2}{a} \left\lceil \frac{N-|S|}{2} \right\rceil-1\\
    &\overset{(b)}{\geq} \frac{|S|}{M} + \frac{K(N-|S|)}{N}-1\\
    & = \frac{|S|(N-KM)+KMN}{NM}-1\\
    &\overset{(c)}{\geq} \frac{N}{M} -1\\  
\end{align*}
And when $|S|\leq M$
\begin{align*}
    K_S(\mathcal{I}_p) &=
     \frac{N - |S|}{M - |S|} - 1 \geq \frac{N}{M}-1 
\end{align*}
Therefore, in both the cases we have
\begin{align*}
    K_S(\mathcal{I}_p)-K_o &\geq \frac{N}{M}-1 - \left(\frac{N}{2M} - 1 + \frac{KN}{b}\right)\\
    &= \frac{N}{2M} - \frac{KN}{b} = \frac{NK}{2}\left(\frac{1}{KM} - \frac{2}{b}\right)
\end{align*}
where $(a)$ follows from observing that $\frac{2}{a}>\frac{2}{b}$ and $\frac{|S^c \cap S_1|}{|S^c|} \geq \frac{1}{2}$ where $S^c = \{ 1, 2, \cdots, N \} \setminus S$. Since $|S^c|=N-|S|$, at least $\frac{N-|S|}{2}$ files of $S^c$ have request probability $\frac{2}{a}$ and the rest have a request probability of at least $\frac{2}{b}$. $(b)$ follows immediately if $N-|S|$ is even. If $N-|S|$ is odd, then $\frac{2}{b} \lfloor \frac{N-|S|}{2} \rfloor + \frac{2}{a} \lceil \frac{N-|S|}{2} \rceil = \frac{N-|S|-1}{2} \cdot \frac{2}{N} + \frac{2}{a} \geq \frac{N-|S|-1}{N} + \frac{1}{a} + \frac{1}{b} = \frac{N-|S|}{N}$. Finally, $(c)$ follows by observing that $|S|(N-KM)$ is minimum when $|S|=N$ since $N-KM<0$ and $0 \leq |S| \leq N$.

A similar argument holds for $\mathcal{I}_q$ as well.
\end{proof}
\subsection{Divergence Calculation}\label{app:ddl}
\begin{lemma}\label{lemma:ddl}
The KL-divergence between the probability measures $\mathbb{P}_p^{\pi}$ and $\mathbb{P}_q^{\pi}$ induced by policy $\pi$ on the set of all possible history sequences $s\in \mathcal{S}$ with a horizon $T$ under instances $\mathcal{I}_p$ and $\mathcal{I}_q$, respectively, is
\[D_{KL}(\mathbb{P}_p^\pi||\mathbb{P}_q^\pi) = TD_{KL}(p||q)\]
where $D_{KL}(p||q)$ is the divergence between the underlying popularity distributions  $p$ and $q$ which is given by
\[D_{KL}(p||q) = \left(\frac{N}{a}-\frac{N}{b}\right) \log\left(\frac{b}{a}\right)\]
\end{lemma}    
\begin{proof}
Consider the interleaved sequence $s$ of observed file requests ($X_i$) and caching decisions ($C_i$) $\{C_1,X_1,C_2 \cdots C_T,X_T\}$ and let $\mathcal{S}$ denote the set of all possible sequences. $\mathbb{P}_p^{\pi}$ and $\mathbb{P}_q^{\pi}$ denote the probability measures on these observed sequences under their respective instances given the policy $\pi$. Also, note that the observed file requests at any point in time are only dependent on the underlying popularity distribution and are independent of whatever is present in the cache. 
\begin{align*}
    \mathbb{P}_p^{\pi}(s) = & \prod_{i= 1}^T \mathbb{P}_p(X_i|C_i,X_{i-1},C_{i-1}\cdots C_1, \pi) \\ & \times\mathbb{P}(C_i|X_{i-1},C_{i-1}\cdots C_1,\pi) 
    \\ & = \prod_{i= 1}^T \mathbb{P}_p(X_i) \times\mathbb{P}(C_i|\mathcal{H}(i),\pi)
\end{align*}
Where $\mathcal{H}(t)$ denotes the history i.e., the set of all cache decisions and request patterns seen up to time $t-1$. Similarly one has
\[\mathbb{P}_q^{\pi}(s)= \prod_{i= 1}^T \mathbb{P}_q(X_i) \times\mathbb{P}(C_i|\mathcal{H}(i),\pi)\]
The divergence $ D_{KL}(\mathbb{P}_p^{\pi}||\mathbb{P}_q^{\pi})$ can now be written as
\begin{align*}
    = & \sum_{s \in \mathcal{S}} \mathbb{P}_p^{\pi}(s)\log\left(\frac{\mathbb{P}_p^{\pi}(s)}{\mathbb{P}_q^{\pi}(s)}\right)\\
    =& \sum_{s \in \mathcal{S}} \mathbb{P}_p^{\pi}(s)\log\left(\frac{\prod\limits_{i= 1}^T \mathbb{P}_p(X_i) \times\mathbb{P}(C_i|\mathcal{H}(i),\pi)}{\prod\limits_{i= 1}^T \mathbb{P}_q(X_i) \times\mathbb{P}(C_i|\mathcal{H}(i),\pi)}\right)\\
    = & \sum_{s \in \mathcal{S}} \mathbb{P}_p^{\pi}(s)\log\left(\frac{\prod\limits_{i= 1}^T \mathbb{P}_p(X_i)}{\prod\limits_{i= 1}^T \mathbb{P}_q(X_i)}\right)\\    
   = &\sum_{s \in \mathcal{S}}\mathbb{P}_p^{\pi}(s)\sum_{i= 1}^T \log\left(\frac{ \mathbb{P}_p(X_i)}{\mathbb{P}_q(X_i)}\right)\\
   =& \sum_{i= 1}^T\sum_{s \in \mathcal{S}} \mathbb{P}_p^{\pi}(X_{i}|s\setminus X_i)\mathbb{P}_p^{\pi}(s\setminus X_{i})\log\left(\frac{ \mathbb{P}_p(X_{i})}{\mathbb{P}_q(X_{i})}\right)\\
   =& \sum_{i= 1}^T\sum_{s \in \mathcal{S}_i} \mathbb{P}_p^{\pi}(s\setminus X_{i})\sum_{X_i}\mathbb{P}_p^{\pi}(X_{i}|s\setminus X_{i})\log\left(\frac{ \mathbb{P}_p(X_{i})}{\mathbb{P}_q(X_{i})}\right)\\  
   =& \sum_{i= 1}^T\sum_{s \in \mathcal{S}_i} \mathbb{P}_p^{\pi}(s\setminus X_{i})\sum_{X_i}\mathbb{P}_p(X_{i})\log\left(\frac{ \mathbb{P}_p(X_{i})}{\mathbb{P}_q(X_{i})}\right)\\ 
   =& \sum_{i= 1}^T\sum_{s \in \mathcal{S}_i} \mathbb{P}_p^{\pi}(s\setminus X_{i})D_{KL}(p||q)\\
   =& \sum_{i=1}^T D_{KL}(p||q) = TD_{KL}(p||q)
\end{align*}
where $\mathcal{S}_i$ is the set of all possible sequences $\{C_1,X_1,C_2 \cdots C_i, C_{i+1}, X_{i+1}, \cdots, C_T, X_T\}$ and $\mathcal{X}_i$ is the set of all possible request vectors (i.e. $[N]^K$). (i.e. $\mathcal{S} = \mathcal{S}_i \times \mathcal{X}_i$)

$D_{KL}(p||q)$ can be written as
\begin{align*}
    D_{KL}(p||q) &= \sum_{i=1}^N p_i\log\left(\frac{p_i}{q_i}\right)\\
    & = \sum_{i=1}^{\frac{N}{2}} \frac{2}{a}\log\left(\frac{b}{a}\right) + \sum_{\frac{N}{2}+1}^N\frac{2}{b}\log\left(\frac{a}{b}\right)\\
    & = \left(\frac{N}{a}-\frac{N}{b}\right) \log\left(\frac{b}{a}\right)
\end{align*}
\end{proof}

\end{document}